\documentclass[12pts]{article}
\usepackage{graphicx}
\include{url}

\title{Analysis of networking characteristics of different personality types}
\author{Charilaos Mylonas \\ ETH Z\"urich \\ Social Network Analysis}

\begin{document}
\maketitle
\newpage
\bibliographystyle{plain}
\section{Introduction}
For many years it has been assumed by psychologists that it is possible to classify an individual according to his personallity. The pioneer in this field was the renowned psychologist C.G.Jung \cite{jungBook} and his contributions led to the new field of study of the so called \textbf{jungian typology}. Regardless of the benefits in prevention and identification of mental illness, typology is used by human resources departments as a guidance for constructing groups with compatible personalities or for personel selecton \cite{hr} and job orientation companies to give personallity compatible recommendations. However there is little definitive quantitative evidence of the existence of psychological types or the validity of the testing procedure. It is the purpose of this study to try and identify if there is any correlation between the social networking behavior of individuals and their personality type, and if the interpretation of their personality type results in intuitively valid observed social networking behavior. The tool used for the classification of the survey participants was a test for the Myers-Briggs Type Indicator (MBTI) \cite{mbti} and a portion of their facebook social network was acquired through the facebook API as it is visible from the authors perspective (connection of common friends).

\section{MBTI basics}

Jungian typology assumes that personality is shaped with an individual's preference to gather information directly with the sensory organs or through reflection and imagination (cognitive functions of \textbf{sensation} and \textbf{intuition} respectively) and the way a person takes decisions (psychic functions of \textbf{feeling} and \textbf{thinking}). Moreover, he classified these functions as rational or \textbf{Judging}(thinking and feeling) or irrational or \textbf{Perceiving} (sensing and intuition). Jung theorized that all people use all four functions but the effect of personality emerges due to the fact that different people seem to prefer to use one of them more often (the dominant or primary function).

The type description is complemented with the so-called \textbf{attitudes}. Jung's definition of an attitude is "readiness of the psyche to react in a certain way"\cite{jungBook}. Myer and Briggs tried to capture the attitude's effect by refering to how a person is getting motivation and "energy". The two main attitudes recognised in this setting is \textbf{Introversion}, who describes the attitude of people with the tendency to feel more comfortable and gain their "energy" when alone and \textbf{Extraversion} which describes the attitude of people who feel more comfortable and "recharge" in social situations. Interestingly enough, the so called "attitude" of an individual is best assesed by examining his social behavior so it should be expected that there is a clear correlation of extraversion and increased social behavior as measured by network metrics. All cognitive functions are supposed to be influenced by the attitude. 

It is quite necessary at this point to describe what the 4 letters of the MBTI mean. The first letter describes the attitude and it is either "E" (\textbf{E}xtrovertion) or "I" (\textbf{I}ntroversion). The second letter describes which is the dominant information gathering function and it is either N (i\textbf{N}tuition) or S (\textbf{S}ensation). The third letter denotes what is the dominant decision making function and it is either T (\textbf{Thinking}) or F (\textbf{F}eeling).

Finally according to the theory of the MBTI people also have a preference in using their perceiving or judging function. Extraverted individuals show their dominant perceiving or judging function according to how they are classified in the last MBTI dimension and introverted individuals show their auxiliary perceiving or judging function. The last letter is either "P" (\textbf{P}erceiving) or "J" (\textbf{J}udging). This means that an individual classified as ENFJ would use the dominant judging function which is feeling in his social interactions where an individual classified as an INFJ would show thinking (the dual function of feeling) in his social interactions. It is quite important to this study to understand that according to the theory supporting the MBTI the general attitude (extraversion/introversion) works in conjuction with the preference to judging or perceiving to shape a person's social behavior.

\begin{figure}
\centering
\includegraphics[width=0.7\textwidth]{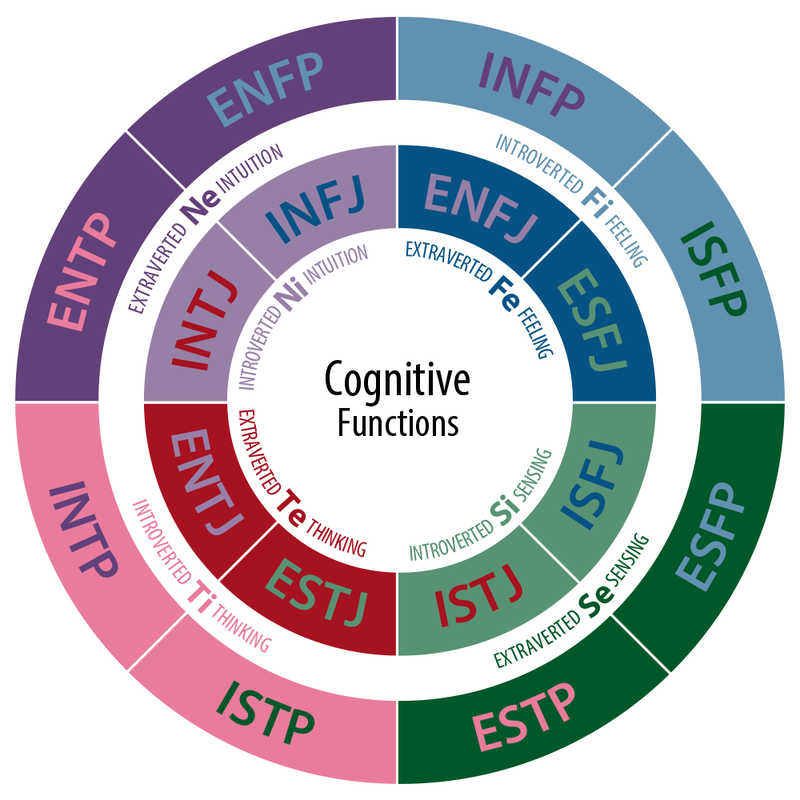}
\label{fig:types}
\caption{"A diagram depicting the cognitive functions of each type. A type's background color represents its Dominant function, and its text color represents its Auxiliary function. A diagram depicting the cognitive functions of each Myers-Briggs personality type. A type's background color represents its dominant function, and its text color represents its auxiliary function." source \cite{wikipediaType}}
\end{figure}

The MBTI result according to the preceding, is a test that classifies different personalities in 16 different types that are summarized in figure \ref{fig:types}

\section{Details on the acquisition of data, identity of the sample}

A free online version of the MBTI test was used \cite{mbtitest}. The results of the test were self reported by the participants in a google spreadsheet. Only 16 participants were involved in the survey possibly due to the large number of questions involved (60 questions). However, it is believed that the participants conducted the test thoroughly and the profile of their personality types seems to match the author's observations. It has to be noted that there are personality tests that have widely replaced the MBTI classification \cite{bigfive} , but they are found to have some correlations with each other \cite{correlations} \cite{correlations2}. For the network data acquisition the \textbf{RFacebook} R package was used and a facebook developer key was acquired. Utility functions were coded in order to perform some manipulations on the network that will be elaborated in the following. The identity of the sample with respect to the personality types is given in table \ref{tab:identity}.

\begin{figure}
\centering
\includegraphics[width=0.7\textwidth]{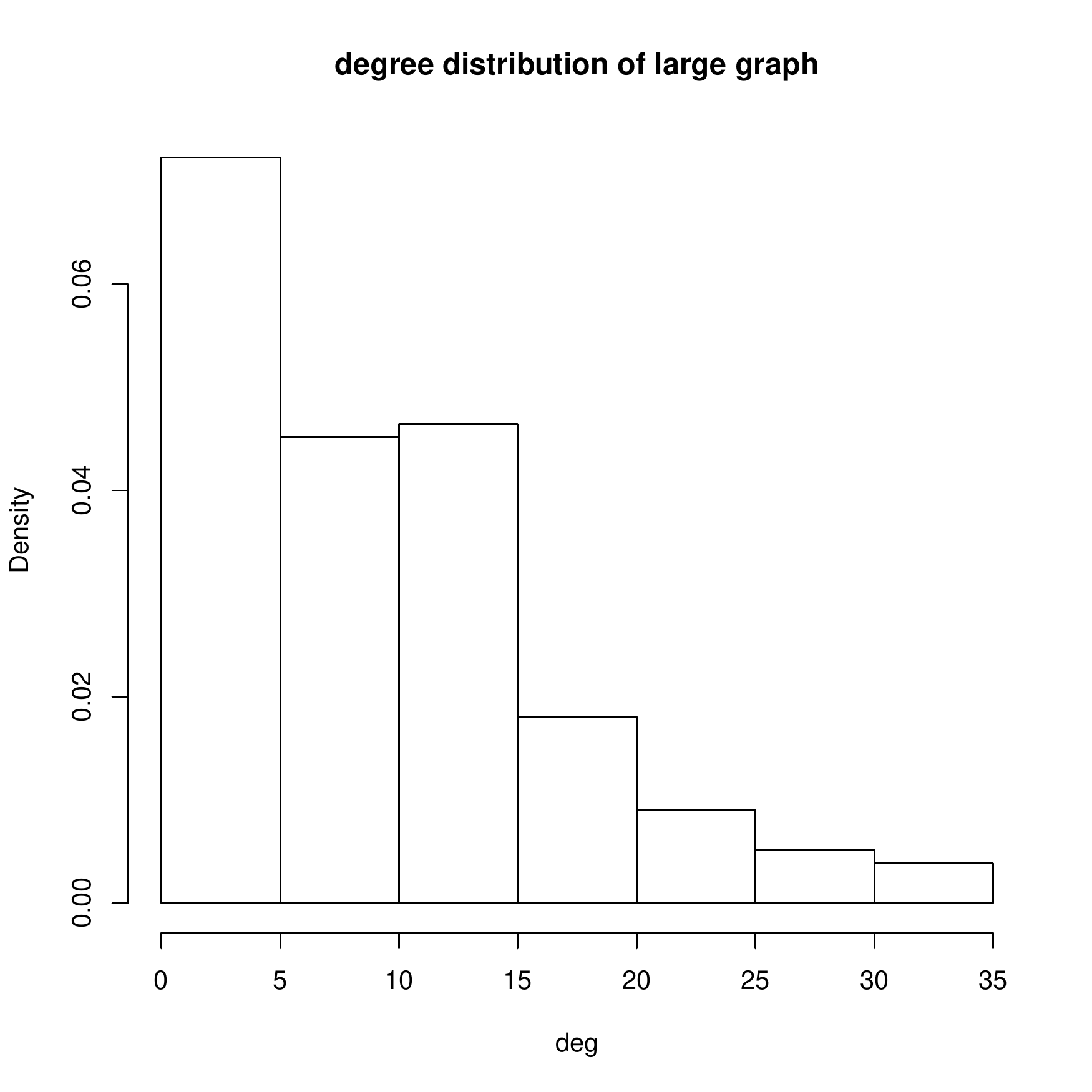}
\caption{The degree distribution of the large network. }
\label{fig:degreeDist}
\end{figure}

\begin{figure}
\centering
\includegraphics[width=0.7\textwidth]{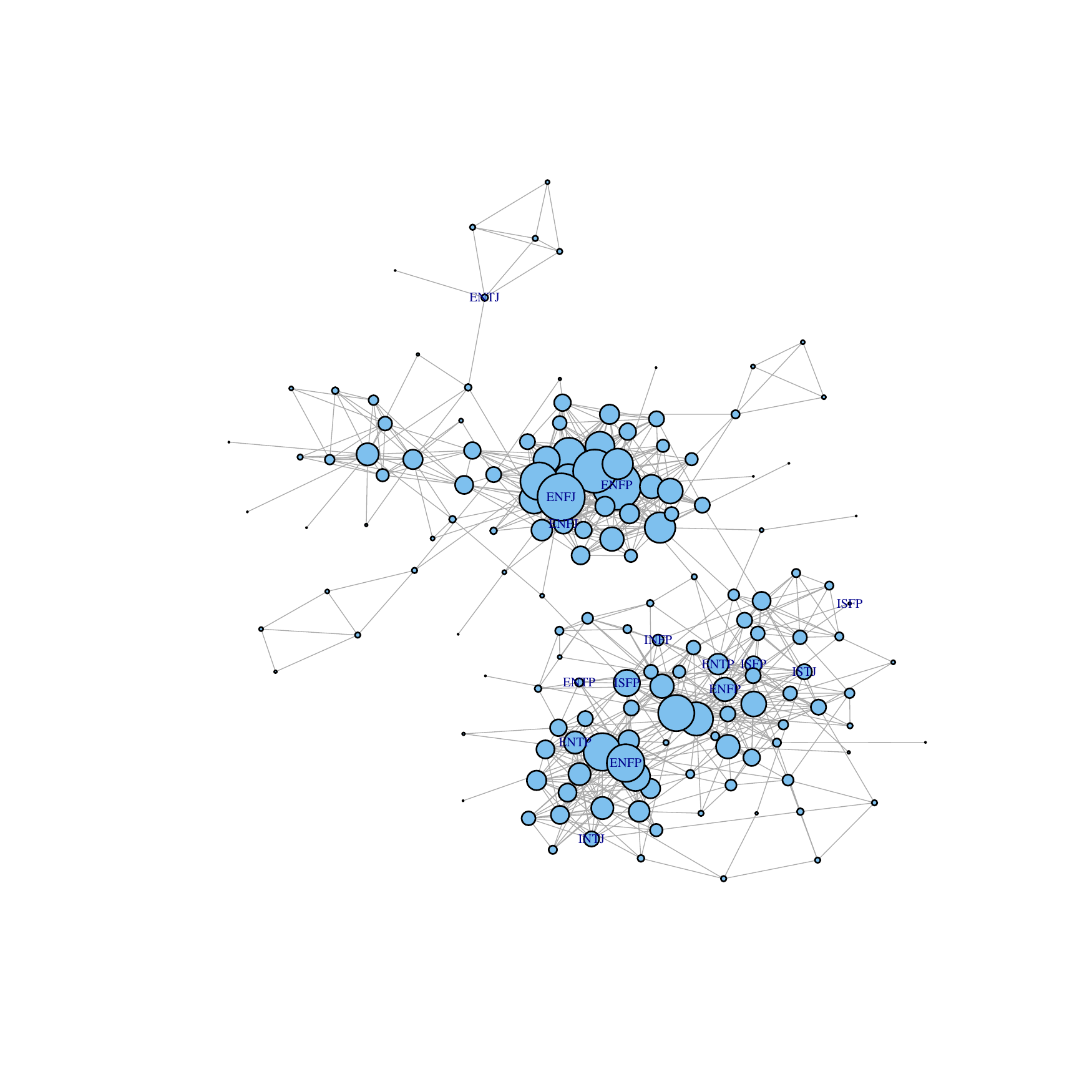}
\caption{The facebook graph studied. The size of the vertices is proportional to the degree centrality measure.}
\label{fig:totalGraph}
\end{figure}

\begin{table}[ht]
\centering
\begin{tabular}{rrrrrrrrr}
  \hline
 &  E & I & S & N & T & F & J & P \\ 
  \hline
\#N &  9 & 6 & 4 & 11 & 6 & 9 & 5 & 10 \\ 
\% & 0.60 & 0.40 & 0.27 & 0.73 & 0.40 & 0.60 & 0.27 & 0.67 \\
\hline
\end{tabular}
\caption{personality dichotomies of the sample}
\label{tab:identity}
\end{table}
\newpage
Singletons were removed from the graph. The resulting connections are summarized in figure \ref{fig:totalGraph}. The degree distribution is given in figure \ref{fig:degreeDist}.

\section{Personality and Social Networking}
Due to the sparcity of the personality type data in the present study, it is quite hard to compute network measures that depend on knowledge of the attributes of many vertices such as assortativity with respect to personality types. Such a study would give interesting insights about the applicability of personality type as a parameter in social network modeling and provide quantitative experimental data as validation for the personality type classification tests that are only theoretically (empirically) grounded. 

\subsection{Centrallity measures and extraversion}
It was found that there is a signifficant correlation of the measure of the Extraversion/Introversion bisection (first letter of the MBTI classifier) and the centrality measure of an individual. Namely, centrality measures for individuals classified as introverted by the MBTI test is strictly lower than the centrality measures for extraverted individuals. There is a high standard deviation but this might be due to some gross missclassifications and other effects. The results for the centrallity are summarized in table \ref{tab:centr}.
\begin{table}[ht]
\centering
\small
\begin{tabular}{|r||rr|rr|rr|rr|}
  \hline
 & E & I & N & S & T & F & J & P \\ 
  \hline
mean(evc) & 0.28 & 0.003 & 0.22 & 0.004 & 0.003 & 0.279 & 0.30 & 0.10 \\ 
  sdev(evc) & 0.43 & 0.003 & 0.40 & 0.003 & 0.003 & 0.431 & 0.43 & 0.32 \\ 
  mean(deg) & 18.78 & 10.500 & 17.09 & 11.000 & 10.667 & 18.66 & 15.0 & 15.7 \\ 
  sdev(deg) & 10.98 & 5.541 & 10.53 & 6.976 & 4.502 & 11.358 & 11.1 & 9.82 \\ 
  mean(btw) & 643.9 & 614.680 & 541.81 &  880.95 & 296.62 & 855.9 & 518.5 & 689.1 \\ 
  sdev(btw) & 675.9 & 1083.90 & 646.81 & 1292.4 & 362.78 & 984.0 & 685.5 & 916.2 \\ 
   \hline
\end{tabular}
\caption{Summary statistics of centrality measures with respect to personality. Legend:  evc = eigenvector centrality as given by the pagerank vector of the full graph, btw = betweenness, deg = degree centrality} 
\label{tab:centr}
\end{table}

\subsection{Characteristics of local networks}
It is intuitively expected that if there is some sort of assortative process relating to social relationships that are influenced by the personalities of the social actors. The influence of interactions of personalities might be masked by other effects as social status, education (and importance an individual assignes to them with respect to social relations), interests, hobies, political views etc but in this study these effects are neglected on grounds that in reallity there is only one degree of separation between the individuals in the social network (the author) so the effects of the afformentioned should be limited.

Initially we turn our attention to the distribution of the minimum path lengths that different personality types present with all the other actors in the network. It is believed that if there is some sort of mechanism of preferential attachment with respect to personalities (and the personality assesment test gives accurate enough results) the distribution of the path lengths should have some speciffic forms with respect to different personalities. The mean and the median of the path length samples were acquired in order to capture the path length distribution skewness. The path lengths are averaged for all the individuals classified with a speciffic attitude (E/I), cognitive function preference (N/S - T/F) and preference to judging or perceiving. It seems that although the type speciffic skewness is always positive,(meaning that there are comparably more people with short paths between them than with large paths between them) and the mean path length is roughly the same for all classifications, there are some differences in the path length distribution skewness. The skewness for the path distribution of different type classifications give the interesting result, (which is quite hard to safely support due to the scarcity of classification data and other obvious shortcomings of the present study) that an ENFP personality, that would present the most positively skewed path length distribution than any other type, would probably be the best connected and have the most acquaintances. Moreover, we would expect individuals classified closer to that personality type to be more active in terms of (internet) social networking. It is tempting to speculate that this might be a statistical effect relating to preferential attachment with respect to personality. In pop-psychology websites \cite{popPsych}, the ENFP type is presented as "{\it{warm, enthusiastic people ... (with) great people skills}} so there seems to be some anecdotal justiffication for the apparent effect of the speciffic personality type as an important factor in social relationships.

\begin{table}[ht]
\centering
\begin{tabular}{|r||rr|rr|rr|rr|}
  \hline
 & E & I & N & S & T & F & J & P \\ 
  \hline
mean(mpl) & 3.42 & 3.44 & 3.46 & 3.35 & 3.75 & 3.22 & 3.70 & 3.30 \\ 
  skn(mpl) & 1.25 & 0.81 & 1.32 & 0.80 & 0.89 & 1.33 & 0.57 & 0.93 \\ 
   \hline
\end{tabular}
\end{table}

\section{Conclusion}
An attempt was made to correlate social network quantitative measures with measures of personality. There were obvious limitations to the current study, the most prominent one being the lack of sufficient personality classifiction data and the fact that only one network of personal acquaintances was studied. Intuitive results were acquired regarding the extraverted attitude and high importance with respect to centrality. A big thank you deserves to Mr. Wehrli for the inspired lecture, the critical comments regarding this work (that unfortunately were not entirely implemented...) and my partners that decided to go their own way and produced another interesting project.
\newpage
\bibliography{bibliography}
\end{document}